\documentclass[12pt]{article}
\usepackage{graphicx}
\newcommand{\lsim}{\,\lower.5ex\hbox{$\stackrel{\textstyle <}{\sim}$}\,}
\begin{document}
\title{The Neutral Decay Modes of the Eta-Meson\footnote{Contribution
to the Eta Physics Handbook, Workshop on Eta Physics, Uppsala 2001}}

\author{B.M.K. Nefkens\footnote{\texttt{nefkens@physics.ucla.edu}} and 
        J.W. Price\footnote{\texttt{price@physics.ucla.edu}}\\
        \emph{University of California at Los Angeles}\\
        \emph{Los Angeles, CA 90095--1547 USA}
       }
\maketitle

\begin{abstract}
The neutral decay modes of the $\eta$ meson are reviewed.  The most
recent results obtained with the Crystal Ball multiphoton detector at
BNL are incorporated.  This includes a new, precise result for the
slope parameter $\alpha$ of the Dalitz plot in $\eta\to 3\pi^0$ decay
and a new, lower branching ratio for $\eta\to\pi^0\gamma\gamma$ which
is consistent with chiral perturbation theory. Recently-obtained
limits are given for novel tests of $CP$ and \emph{C} invariance based
on several rare $\eta$ decays.
\end{abstract}

\section{Introduction}
The $\eta$ meson has the interesting feature that all its possible
strong decays are forbidden in lowest order: $\eta\not\to2\pi$ and
$\eta\not\to4\pi^0$ by $P$ and $CP$ invariance,
$\eta\not\to3\pi$ because of $G$-parity conservation as well as
isospin and charge symmetry invariance.  First order electromagnetic
$\eta$ decays are forbidden as well: $\eta\to\pi^0\gamma$ by
conservation of angular momentum, $\eta\to2\pi^0\gamma$ and
$\eta\to3\pi^0\gamma$ by \emph{C} invariance.  Only
$\eta\to\pi^+\pi^-\gamma$ occurs, but at a suppressed rate because it
involves the anomaly.  The first allowed decay is the second-order
electromagnetic transition $\eta\to2\gamma$. 

The width of the $\eta$ is about 1.3 keV; this is 5 orders of
magnitude smaller than a typical strong decay, such as the $\rho$
meson.  This feature makes $\eta$ decays $10^5$ times more sensitive
than $\rho$ or $\omega$ decays at a comparable branching ratio for
testing invariances. 

The physical $\eta$ is a mixture of the pseudoscalar SU(3) octet
($\eta_8$) and singlet ($\eta_0$) parametrized by the mixing angle
$\theta$:
\begin{eqnarray*}
|\eta\rangle & = & \cos\theta|\eta_8\rangle - 
                   \sin\theta|\eta_0\rangle \\
  & = & \frac{1}{6}\sqrt{6}\cos\theta
          |\overline{u}u + \overline{d}d - 2\overline{s}s\rangle - 
        \frac{1}{3}\sqrt{3}\sin\theta
          |\overline{u}u + \overline{d}d + \overline{s}s\rangle
\end{eqnarray*}
The value for $\theta$ is $(20\pm2)^\circ$.  If we choose
$\theta=19.5^\circ$, we have the following remarkable makeup of the
physical $\eta$:
\[
|\eta\rangle = 
  \frac{1}{3}\sqrt{3}|\overline{u}u + 
                      \overline{d}d - 
                      \overline{s}s\rangle,
\]
which means that the $\eta$ is an eigenstate of the $I$, $U$, and $V$
operators of SU(3).  We also have $|\eta'\rangle =
\sin\theta|\eta_8\rangle + \cos\theta|\eta_0\rangle =
\frac{1}{6}\sqrt{6}|\overline{u}u + \overline{d}d +
2\overline{s}s\rangle$.  In the limit $u=d\equiv q$, we have 
$\eta'=\frac{1}{3}\sqrt{6}|\overline{q}q + \overline{s}s\rangle$.
The $\eta'$ is an eigenstate of the operator that interchanges the $s$
and the $q$ quarks. 

There are more than 14 different neutral $\eta$ decays (see
Table~\ref{tab:alldecays} at the end).  They make up (71.6 $\pm$
0.4)\% of all $\eta$ decays~\cite{PDG01}.

\section{$\eta\to2\gamma$}
The largest branching ratio of the $\eta$ is into two photons:
\begin{equation} 
BR(\eta\to2\gamma) = (39.5\pm0.2\pm0.3)\%.
\end{equation}
It was measured in a dedicated experiment~\cite{Abe96} carried out at
SACLAY with the SPES II $\eta$  ``factory" using the $p d\to
^3$He$\eta$ production reaction.

\begin{figure}
\includegraphics[width=\textwidth]{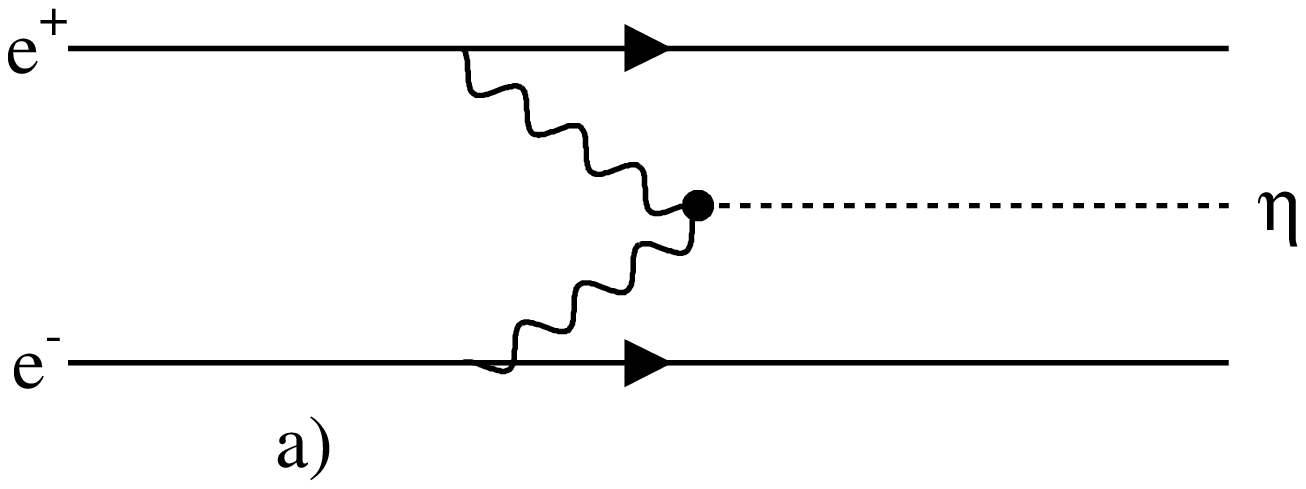}
\includegraphics[width=\textwidth]{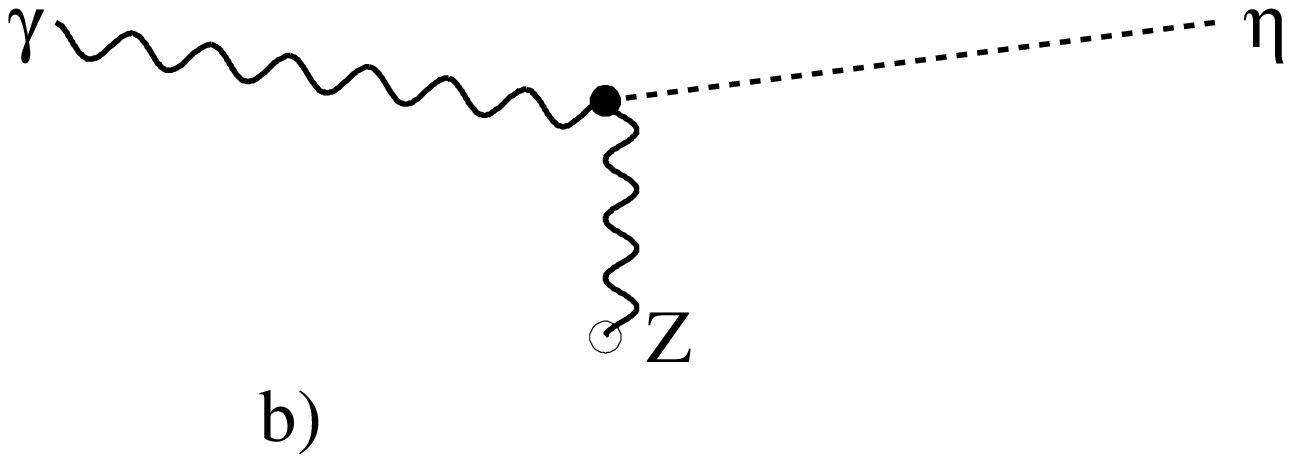}
\caption{\label{fig:twogamma}\small Feynman diagrams of $\eta$
production for the determination of $\Gamma(\eta\to2\gamma)$.  a)
$\eta$ production in QED.  b) $\eta$ production by the Primakoff
effect.} 
\end{figure}

The decay rate $\Gamma(\eta\to2\gamma)$ has been determined by two
means. The first one is the QED process 
$e^+e^- \to e^+e^-\gamma^*\gamma^* \to e^+e^-\eta \to e^+e^-2\gamma$
(see Fig.~\ref{fig:twogamma}a). 
The calculation of the rate is believed to be well-understood and the
uncertainty due to the virtual photon form factor is
small~\cite{PDG94note}.  The results of four
experiments~\cite{Bar85,Wil88,Roe90,Bar90} are given in
Table~\ref{tab:twogamma}.  They are mutually consistent; the average
value is 
\begin{equation}
\Gamma(\eta\to2\gamma) = 0.510\pm0.026\ \mathrm{keV}.
\label{eq:etaqed}
\end{equation}

\begin{table}
\caption{\label{tab:twogamma}\small Experimental results for
$\Gamma(\eta\to2\gamma)$ measured in
$e^+e^-\to e^+e^-\eta$.}
\begin{center}
\begin{tabular}{lcc}\hline\hline
Experiment & Reference & $\Gamma(\eta\to2\gamma)$\\
\hline
Bartel \emph{et al.}   & \cite{Bar85} & $ 0.53  \pm 0.04  \pm 0.04$  \\
Williams \emph{et al.} & \cite{Wil88} & $ 0.514 \pm 0.017 \pm 0.035$ \\
Roe \emph{et al.}      & \cite{Roe90} & $ 0.490 \pm 0.010 \pm 0.048$ \\
Baru \emph{et al.}     & \cite{Bar90} & $ 0.51  \pm 0.12  \pm 0.05$  \\
\hline\hline
\end{tabular}
\end{center}
\end{table}
There are 2 older measurements of very limited
statistics~\cite{Wei83,Aih86} which we are not using.  Finally, there
is a different type of measurement based on the Primakoff
effect~\cite{Bro74}, $\gamma A\to\gamma\gamma A$ (see
Fig.~\ref{fig:twogamma}b), which gives
\begin{equation}
\Gamma(\eta\to2\gamma) = 0.324 \pm 0.046\ \mathrm{keV}.
\end{equation}
This experiment suffers from uncertainties in the dependence of the
Primakoff form factor on the momentum transfer and production angle,
and on the systematic error in the phase of the interference term.
Because the difference between the Primakoff and the QED process is
$4\sigma$ and the four QED measurements are consistent with one
another and have fewer theoretical uncertainties, we recommend using
the value of Eq.~2 rather than the average of both methods as done by
the Particle Data Group~\cite{PDG01}.  Combining Eqs.~1 and 2, we
obtain for the total $\eta$ decay rate 
\begin{equation}
\label{eq:etawidth}
\Gamma(\eta\to all) = 1.29 \pm 0.07\ \mathrm {keV},
\end{equation}
which is 11\% larger than the PDG value.

\begin{figure}
\includegraphics[width=\textwidth]{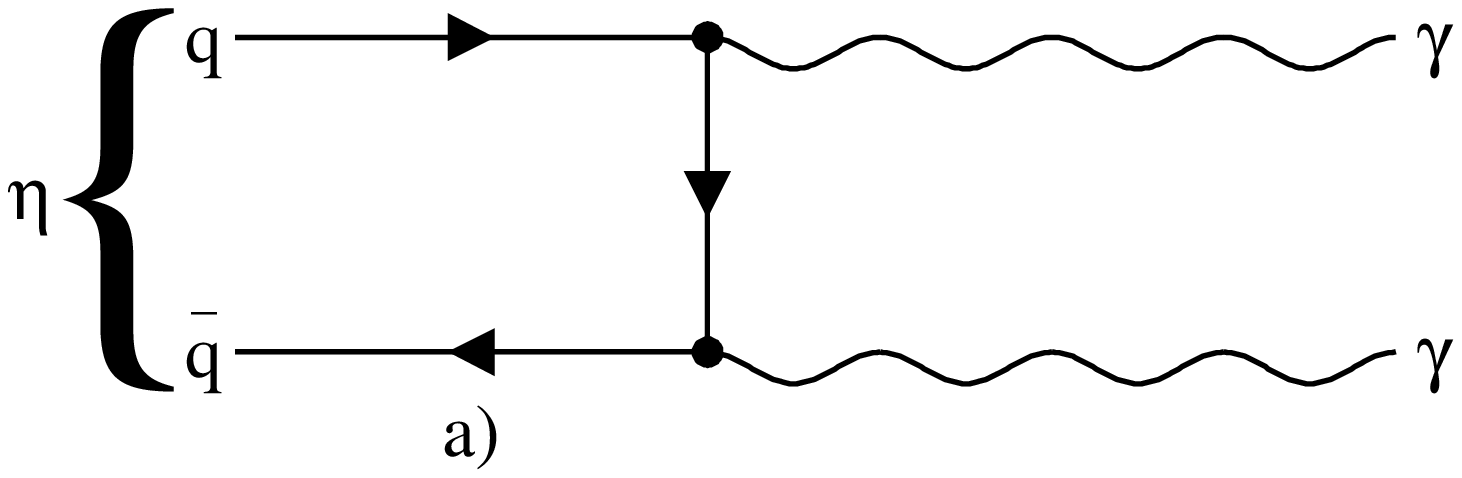}
\includegraphics[width=\textwidth]{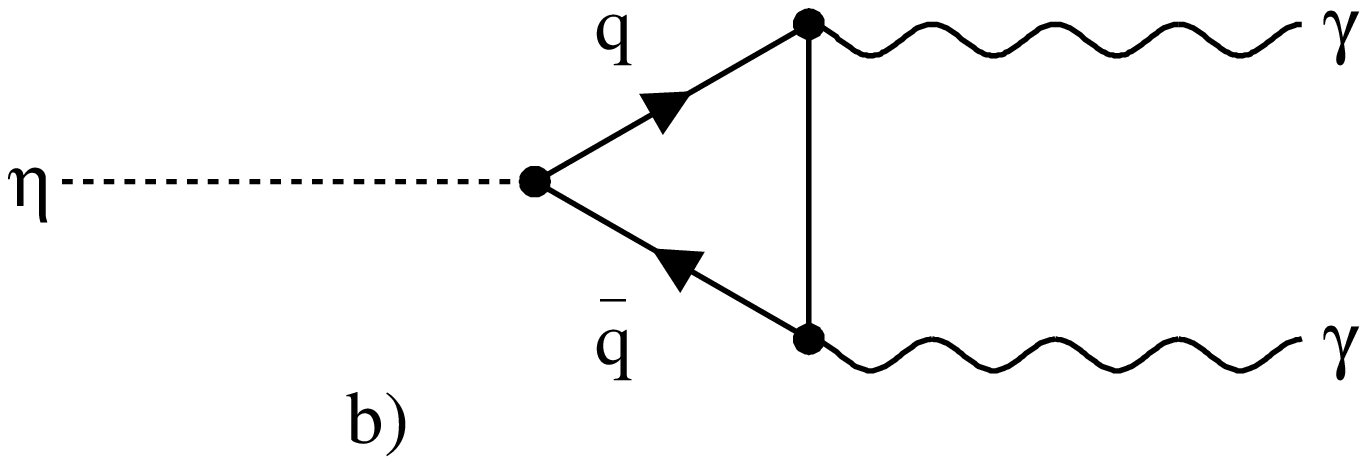}
\caption{\label{fig:etadecay}\small Feynman diagrams for the decay
$\eta\to\gamma\gamma$. a) Electromagnetic annihilation of a
constituent quark with its antiquark. b) Triangle graph incorporating
the QCD anomaly.} 
\end{figure}

$\Gamma(\eta\to 2\gamma)$ can be calculated in different ways.  The
order of magnitude is readily obtained by scaling the positronium
decay rate treating $\eta\to2 \gamma$ as the electromagnetic
annihilation of a constituent quark and its antiquark (see
Fig.~\ref{fig:etadecay}a), thus
\begin{equation}
\Gamma(\eta\to2\gamma) = 
  \left(\frac{m(\eta)}{m(e^+e^-)}\right)^3
  \Gamma(\mathit{positr.}\to2\gamma) = 
  0.81\ \mathrm{keV},
\end{equation}
which agrees to a factor of 1.6 with the measured value.  For
comparison, note that this scaling law gives
$\Gamma(\pi^0\to2\gamma)=12$ eV, which is 1.5 times the measured
result.  This calculation gives $\Gamma(\eta'\to2\gamma)=4.3$ keV,
also in good agreement with the measurement. 

It was a stunning
surprise when Sutherland~\cite{Sut67} and Veltman~\cite{Vel67} showed
that in the limit of massless quarks, the 2$\gamma$ decay of $\pi^0$
and $\eta$ is forbidden.  This embarrassing
situation --- after all, the dominant decay of both the $\pi^0$ and
the $\eta$ is into 2 photons --- turned into a triumph when
Adler~\cite{Adl69} and Bell and Jackiw~\cite{Bel69} figured out how
all is saved by the QCD anomaly.  The triangle graph (see
Fig.~\ref{fig:etadecay}b) results in the prediction 
\begin{equation}
\Gamma(\eta\to2\gamma) = 
  \Gamma(\pi^0\to2\gamma)\left(\frac{m_\eta}{m_\pi}\right)^3\times\phi,
\end{equation}
where
\[
\phi = 
\sqrt{18}\left[\frac{F_\pi}{F_8}
                 \cos\theta(Q_u^2 + Q_d^2 - 2Q_s^2) - 
               \sqrt{2}\frac{F_\pi}{F_0}
                 \sin\theta(Q_u^2 + Q_d^2 + Q_s^2)
         \right]^2,
\]
\[
\Gamma(\pi^0\to\gamma\gamma) = 
   \alpha^2m^3_\pi N_c^2(Q_u^2 - Q_d^2)^2/32\pi^3F_\pi^2.
\]
The last equation gives a value of 7.6 eV for
$\Gamma(\pi^0\to\gamma\gamma)$, in agreement with the measurement of
($7.9\pm0.7$) eV.  Using $F_8=1.3F_\pi$, $F_0=(1.05\pm0.04)F_\pi$, and
$\theta=19.5^\circ$, we obtain $\phi=1.3\pm0.2$ and
$\Gamma(\eta\to2\gamma)=0.66$ keV, which is close to the experimental
value (Eq.~\ref{eq:etaqed}).  A more sophisticated handling of the
octet-singlet mixing that involves two mixing angles is available in
the literature~\cite{Fel00}.

These results have been hailed as experimental verification that the
number of colors, $N_c$, is 3.  On detailed inspection this is not
quite so as discussed in Ref.~\cite{Bar01}.  The branching ratio for
$\eta\to2\gamma$ plays a pivotal role in experimental eta physics, as
it is used in the determination of the $\eta$ flux in many
experiments.  It appears prudent therefore to have a new measurement
of $BR(\eta\to2\gamma)$, preferably with a $4\pi$ acceptance photon
detector. 

\section{$\eta\to3\pi^0$}
The second largest $\eta$ decay mode is $\eta\to 3\pi^0$.  The PDG
recommends the value $BR(\eta\to3\pi^0)=(32.24\pm0.29)\%$.  This value
was determined principally by GAMS-2000~\cite{Ald84} who reported that
$\Gamma(2\gamma)/\Gamma(neutrals) = 0.549\pm0.004$ and
$\Gamma(3\pi^0)/\Gamma(neutrals) = 0.450\pm0.004$, resulting in
$\Gamma(3\pi^0)/\Gamma(2\gamma) = 0.820\pm0.009.$  This agrees with
the average of 3 direct measurements; the PDG quotes $0.825\pm 0.011$
for the ratio.

The decay mode of the $\eta$ into 3 pions violates isospin invariance.
This $\eta$ decay occurs because of the $u$-$d$ quark mass difference,
which can be seen as follows.  The Lagrangian of QCD,
$\mathcal{L}_{QCD}$, can be divided into two parts: 
\[
\mathcal{L}_{QCD} = 
   \mathcal{L}_0+\mathcal{L}_m,
\]
where
\[
\mathcal{L}_0 = 
   -\frac{1}{4}F_{\mu\nu}^{(a)}F^{(a)\mu\nu} + 
   i\sum_q\overline{\psi}^i_q\gamma^\mu(D_\mu)_{ij}\psi^j_q,
\]
\[
F_{\mu\nu}^{(a)} = 
   \partial_\mu A_\nu^a - 
   \partial_\nu A_\mu^a + 
   g_sf_{abc}A_\mu^bA_\nu^c,
\]
\[
(D_\mu)_{ij} = 
   \delta_{ij}\partial_\mu - ig_s\sum_a\frac{\lambda_{i,j}^a}{2}A_\mu^a,
\]
$A$ is the gluon field, $\psi_q$ is the quark field, $g_s$ is the strong
coupling factor, and $f_{abc}$ is the SU(3) structure constant.  Thus,
$\mathcal{L}_0$ depends only on the quark and gluon fields and their
derivatives, and is the same for all quarks.  This is called the flavor
symmetry of (massless) QCD, and isospin symmetry is flavor symmetry
applied to the $u$ and $d$ quarks.  The flavor symmetry is broken by
the quark-mass term, 
\[
\mathcal{L}_m = 
   -\sum_qm_q\overline{\psi}^i_q\psi_{q_i}.
\]
Thus we obtain the following expression for $\pi^0$-$\eta$ mixing:
\begin{eqnarray}
\langle \pi | H_m | \eta \rangle 
  & = & \left\langle \frac{1}{2}\sqrt{2}(\bar{u}u-\bar{d}d) |
             \bar{u}m_uu + \bar{d}m_dd |
             \frac{1}{3}\sqrt{3}(\bar{u}u+\bar{d}d-\bar{s}s)
\right\rangle\\
  & = &\frac{1}{6} \sqrt{6} (m_u - m_d),\nonumber
\end{eqnarray}
where we have used the value $\theta = -19.5^\circ$ for the SU(3)
octet-singlet mixing angle.  $\eta \to 3\pi$ is \emph{not} an
electromagnetic decay as is sometimes stated in the older literature;
it is a \emph{limited} strong decay, which depends only on
$\mathcal{L}_m$ and not on $\mathcal{L}_0$.  We also have $\Gamma
(\eta \to 3 \pi)\sim (m_u - m_d)^2$; it is one of the best ways to
determine the $u$-$d$ quark mass difference.  A related way is via the
ratio
$r=\Gamma(\eta\to3\pi^0)/\Gamma(\eta\to\pi^+\pi^-\pi^0)$~\cite{Ani96}. 

\begin{figure}
\centerline{
\includegraphics[width=9cm]{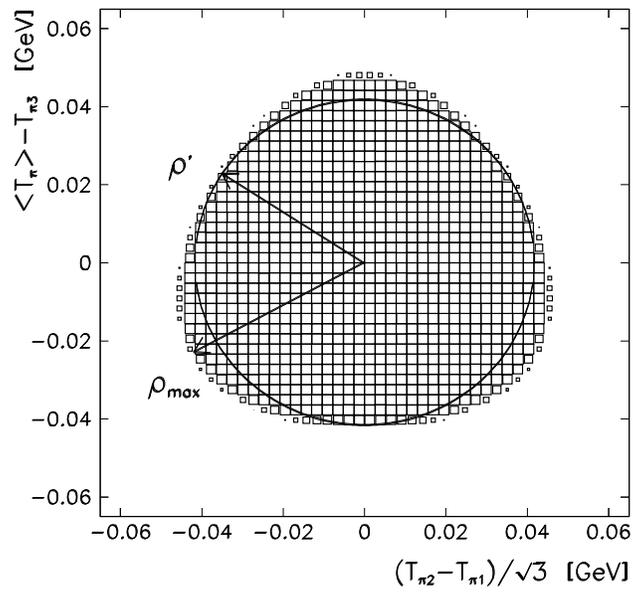}}
\caption{\label{fig:symdalitz}\small Symmetrized Dalitz plot for the
decay $\eta\to3\pi^0$.  The plot is uniform along concentric
circles. The deviation from a circular shape comes from using
relativistic kinematics.} 
\end{figure}

An interesting parameter to investigate is the slope in the Dalitz
plot for $\eta \to 3 \pi^0$.  In lowest order the Dalitz plot should
be uniform because of the identical final state particles.  However,
the $\pi - \pi$ interaction is strong and strongly energy dependent,
which results in a tiny nonuniformity of the Dalitz plot.  To
incorporate the fact that there are 3 identical $\pi^0$'s in the final
state, we use a symmetrized Dalitz plot as illustrated in
Fig.~\ref{fig:symdalitz}.
The deviation of the shape from a circle is the result of the use of
relativistic kinematics.  The experimental Dalitz plot density should
be uniform in concentric circles around the center.  A suitable
variable to transform from the two-dimensional to a one-dimensional
distribution is $z$: 
\[
z = 
   6\sum_{i=1}^3(E_i-m_\eta/3)^2/(m_\eta- m_{\pi^0})^2 = 
   \rho^2/\rho^2_{max},
\]
where $E_i$ is the energy of the $i$th pion in the $\eta$ rest frame
and $\rho$ is the distance from the center of the Dalitz plot.  The
variable $z$ varies from $z=0$, when all three $\pi^0$'s have the same
energy of $m_{\eta} /3$, to $z = 1$, when one $\pi^0$ is at rest. 

The most accurate result comes from the very recent measurement by the
Crystal Ball Collaboration~\cite{Tip01} yielding $\alpha = -0.031
(4)$.  The world data is summarized in Table~\ref{tab:alpha}.  
\begin{table}
\caption{\label{tab:alpha}\small Experimental measurements of the
slope parameter $\alpha$ of the decay $\eta\to\pi^0\pi^0\pi^0$.}
\begin{center}
\begin{tabular}{lcc}\hline\hline
Experiment     & Reference    & $\alpha$\\
\hline
Baglin \emph{et al.}  & \cite{Bag70} & $-0.32  \pm 0.37$            \\
GAMS-2000      & \cite{Ald84} & $-0.022 \pm 0.023$           \\
Crystal Barrel & \cite{Abe98} & $-0.052 \pm 0.017 \pm 0.010$ \\
SND            & \cite{Ach01} & $-0.010 \pm 0.021 \pm 0.010$ \\
Crystal Ball   & \cite{Tip01} & $-0.031 \pm 0.004$           \\
\hline\hline
\end{tabular}
\end{center}
\end{table}
Predictions based on chiral perturbation theory made by Kambor
\emph{et al.}~\cite{Kam96}, who used a dispersion calculation in which
rescattering effects are treated to all orders, give values for
$\alpha$ in the range $-(0.014-0.007)$.  There is agreement with
experiment in the sign but not in the magnitude.
B. Holstein~\cite{Hol01} has suggested the addition of a new dynamical
input. 

For comparison note that the slope parameter in $K_L$ decay is
$2\alpha=(-6.1\pm0.9\pm0.5)\times10^{-3}$~\cite{Lai01}; it agrees in
sign as expected but it is much smaller, presumably because there is
$\frac{1}{3}$ less energy released than in $\eta\to3\pi^0$ decay.  It
will be of interest to make a slope measurement of
$\eta^{\prime}\to3\pi^0$, where the energy release is sufficiently
large to see the effects of a major $\pi - \pi$ resonance. 

\section{\label{sec:pi0gg}$\eta\to\pi^0\gamma\gamma$}
The smallest measured branching ratio of the neutral $\eta$
decays is of the doubly radiative transition $\eta \to \pi^0 \gamma
\gamma$.  At first glance one might expect the BR to be just slightly
smaller than $\eta \to 2 \gamma$ because of the reduced phase space.
In reality the decay rate is suppressed because of the chiral symmetry
of the $\mathcal{L}_0$ term of $\mathcal{L}_{QCD}$.  In terms of the
momentum expansion used in chiral perturbation theory ($\chi$PTh), the
leading $O (p^2)$ term is absent for massless quarks; the $O(p^4)$
term is very small because there is no direct photon coupling to the
$\pi^0$ and $\eta$.  Thus, the decay $\eta \to \pi^0 \gamma\gamma$ is
a (unique) test of the $O(p^6)$ terms of $\chi$PTh.  Various
theoretical estimates are given in Table~\ref{tab:pi0gg}.  
\begin{table}
\caption{\label{tab:pi0gg}\small Theoretical predictions of the decay
$\eta\to\pi^0\gamma\gamma$.} 
\begin{center}
\begin{tabular}{lcc}\hline\hline
Prediction          & Reference       & 
$\Gamma(\eta\to\pi^0\gamma\gamma)$ (eV) \\
\hline
Ko ($O(p^4)$)       & \cite{Ko95}     & $0.004$ \\
Ko                  & \cite{Ko93a}    & $0.47\pm0.20$ \\
Ametller \emph{et al.}     & \cite{Ame92}    & $0.42\pm0.20$ \\
Nemoto \emph{et al.}       & \cite{Nem96}    & $0.92$ \\
Bellucci and Bruno  & \cite{Bel95}    & $0.58\pm0.3$ \\
Ng and Peters (VMD) & \cite{Ng92}     & $0.30^{+0.16}_{-0.13}$ \\
Ng and Peters (Box) & \cite{Ng93}     & $0.70$ \\
\hline\hline
\end{tabular}
\end{center}
\end{table}
The only sufficiently sensitive published result is by
GAMS-2000~\cite{Ald84}. Their result,
$BR(\eta\to\pi^0\gamma\gamma)=7.1\pm1.7\times 10^{-4}$, is based on a
signal of 38 events. It is larger than every prediction based on
$\chi$PTh.  The preliminary result of the CB
experiment~\cite{Pra01,Pra01c} based on 500 events is
$BR(\eta\to\pi^0\gamma\gamma)=(3.2\pm0.9)\times10^{-4}.$  Together
with Eq.~\ref{eq:etawidth}, this yields
$\Gamma(\eta\to\pi^0\gamma\gamma)=0.42\pm0.14\ \mathrm{eV}$, which is
well within the range of the theoretical predictions.  It will be of
considerable interest to measure the Dalitz plot of this decay.
Again, a comparison with $K_L \to\pi^0\gamma\gamma$ is interesting. 

\section{$\eta\to\pi^0\pi^0\gamma\gamma$}
The two pion, doubly radiative decay, $\eta\to\pi^0\pi^0\gamma\gamma$,
is driven by the Wess-Zumino anomaly, and therefore is
expected~\cite{Ko93} to have a very small branching ratio.  It
provides another, though difficult case for testing the higher order
terms in $\chi$PTh.  The original interest in
$\eta\to\pi^0\pi^0\gamma\gamma$ was focused on providing important
limits on the coupling of new, relatively light neutral gauge bosons,
useful already at the level of $BR\lsim10^{-5}$, as discussed e.g.\ by
D. Wyler~\cite{Wyl90}.  The decay amplitude can be divided into two
components: 
\begin{equation}
A(\eta\to\pi^0\pi^0\gamma\gamma) = A_R \pm A_{NR}.
\end{equation}
The first component, $A_R$, is of the resonance-type.  It is
proportional to the on-shell amplitudes for $\eta\to3\pi^0$ and
$\pi^0\to2\gamma$.  Besides being maximized at the kinematics for
$\eta\to3\pi^0$, it has a limited value over the entire phase
space. The non-resonant amplitude, $A_{NR}$, can be evaluated in
$\chi$PTh~\cite{Bel97,Ame97}.  For $s_{\gamma\gamma}>0.23m^2_\eta$,
where $s_{\gamma\gamma}$ is the square of the $\gamma\gamma$ invariant
mass, the non-resonant contribution is larger than the resonance by
more than a factor of two~\cite{Bel97,Ame97}. For an energy cut $\delta
m$ of 20 MeV, a tree-level analysis~\cite{Kno96} gives
\begin{equation}
BR(\eta\to\pi^0\pi^0\gamma\gamma) \simeq 6\times10^{-7}.
\end{equation}
Extending the calculation to one loop~\cite{Bel97}, one obtains
\begin{equation}
BR(\eta\to\pi^0\pi^0\gamma\gamma) \simeq 8\times10^{-8}.
\end{equation}
Experimentally it is very difficult to measure the decay
$\eta\to\pi^0\pi^0\gamma\gamma$ because of the background coming from 
$\eta\to\pi^0\pi^0\pi^0$.  The first, very modest, upper limit has
been obtained very recently by the Crystal Ball
Collaboration~\cite{Pra00}:  
\begin{equation}
BR(\eta\to\pi^0\pi^0\gamma\gamma) < 3.1\times10^{-3}
\end{equation}
with a CL of 90\%.

\section{\label{sec:4gamma}$\eta\to4\gamma$}
The four photon decay of the neutral pseudoscalar mesons is allowed,
and like the two photon decay, is driven by the anomaly described
by the Wess-Zumino-Witten term~\cite{Wes71}.  To this should be added
the $O(p^2)$ terms in the non-anomalous sector~\cite{Lia98}.  Based on
the coupling strength alone, one has
$\Gamma(\eta\to4\gamma)/\Gamma(\eta\to2\gamma) < (\alpha/\pi)^2\simeq
10^{-5}$.  In realistic models the decay $\eta\to4\gamma$ is going to
be much smaller because the lowest orbital angular momentum states are
forbidden by the requirements of gauge invariance and
PCAC~\cite{Sch72}.  The centrifugal barriers of the higher states
inhibit the $\eta\to4\gamma$ decay by extra powers of $(kR)$ where $k$
is the photon momentum and $R$ is the interaction radius.  Thus far no
theoretical estimate or experimental search has been reported. 

For the related case of $\pi^0\to4\gamma$ there are some estimates.  A
Vector-Meson-Dominance-with-PCAC calculation~\cite{Sch72} gives
$BR(\pi^0\to4\gamma)\sim10^{-16}$, while a $\chi$PTh
evaluation~\cite{Lia98} yields $7.1\times 10^{-18}$.  The purely
electromagnetic photon-splitting term alone yields
$BR(\pi^0\to4\gamma)\simeq3\times10^{-11}$ (see Ref.~\cite{Bra95}).  The
experimental upper limit~\cite{McD88} is $BR(\pi^0\to4\gamma)\ <
2\times10^{-8}$ at 90\% CL.

\section{Search for New Physics}
The Standard Model (SM) of the electroweak interactions has been
phenomenally successful in giving a quantitative account of the
various electroweak interactions.  There is no evidence thus far for
any failure.  Yet the SM is not considered to be a theory.  It needs
17 input parameters, not counting the neutrino sector.  It does not
explain such basic features as the existence of three families of
fundamental fermions, the absence of the charge conjugates of the
left-handed neutrinos and right-handed antineutrinos, the generation
of widely different masses of the charged leptons, the origin of \emph
{CP} violation, etc.  It is generally expected that the SM breaks down
somewhere.  A good place to look for New Physics appears to be the
limit of validity of the basic symmetries of charge conjugation $(C)$,
parity $(P)$, and time reversal $(T)$, as well as $CP$ and $CPT$ in
the different interactions. 

If the $CPT$ theorem holds, there are four distinct classes of the
violations of $C,P$, and $T$ (see Table \ref{tab:violations}).  
\begin{table}
\caption{\label{tab:violations}\small The four classes of $C$, $P$,
and $T$ violations assuming $CPT$ invariance.}
\begin{center}
\begin{tabular}{ccc}
\hline\hline
Class & Violated                        & Valid     \\
\hline
1     & $C$, $P$, $CT$, $PT$            & $T$, $CP$ \\
2     & $C$, $P$, $T$, $CP$, $CT$, $PT$ &           \\
3     & $P$, $T$, $CP$, $CT$            & $C$, $PT$ \\
4     & $C$, $T$, $CP$, $PT$            & $P$, $CT$ \\
\hline \hline
\end{tabular}
\end{center}
\end{table}
A particularly stringent limit on simultaneous $P$ and $T$
violation, class 3 of table~\ref{tab:violations}, is obtained from the
smallness of the limit on the electric dipole moment of the neutron.
Yet, all classes, even $\#$3, need better limits, in particular of
electrostrong interactions of the quarks.  If $CPT$ is not valid
there are 7 main classes of $C$, $P$, $T$, and $CPT$ violations (see
Table \ref{tab:moreviolations}). 
\begin{table}
\caption{\label{tab:moreviolations}\small The seven classes of $C$,
$P$, and $T$ violations when $CPT$ is not valid.}
\begin{center}
\begin{tabular}{ccc}
\hline\hline
Class & Violated                          & Valid            \\
\hline
1     & $C$, $CP$, $CT$, $CPT$            & $P$, $T$, $PT$   \\
2     & $P$, $CP$, $PT$, $CPT$            & $C$, $T$, $CT$   \\
3     & $T$, $CT$, $PT$, $CPT$            & $C$, $P$, $CP$   \\
4     & $C$, $P$, $CP$, $CT$, $PT$, $CPT$ & $T$        \\
5     & $C$, $T$, $CP$, $CT$, $PT$, $CPT$ & $P$        \\
6     & $P$, $T$, $CP$, $CT$, $PT$, $CPT$ & $C$        \\
7     & $C$, $P$, $T$, $CPT$        & \\
\hline \hline
\end{tabular}
\end{center}
\end{table}

$CPT$ invariance is a fundamental theorem in quantum field
theory, which results from locality and Lorentz invariance.  Its
validity in the context of quantum gravity is questionable~\cite{Ell93}.
String theory is intrinsically non-local, which could lead to a
violation of $CPT$.  Instantons give rise to charge-non-conserving
transitions on the world sheet, and hence to $CPT$ violations.  Thus,
the extent of the validity of $CPT$ invariance must rest on
experimental evidence which is either dynamic, such as in $K^0$ and
$\eta$ decay, or static, such as the equality of masses, half lives,
magnetic moments, etc., of particles and their antiparticles.  The
most precise static test is $|m_{K^0} -
m_{\overline{K^0}}\vert/m(\mathrm{ave.})<10^{-18}~\cite{PDG01}$.
Testing \emph{CPT} is extremely difficult in dynamic cases.  The decay
$\eta\to\pi^{0}\mu^+\mu^-$ provides a special opportunity. The decay
spectrum must be even in $\cos\phi$, where $\phi$ is the angle between
the $\pi^0$ and the $\mu^{+}$ in the rest frame of the muon
pair~\cite{Nef93,Pai68}. 

\section{$CP$ Violation}
$CP$ symmetry means that the interaction of a set of left-handed
particles is identical to the interaction of the complimentary set of
right-handed antiparticles.  The discovery of a 0.2\% $CP$
violation in 1964 came as a great surprise.  At the time of the
discovery there were no theoretical models for $CP$ violations,
and the experimental upper limit for $K_L\to2\pi$ was 0.3\%!  The
origin of $CP$ violation is still a mystery.  There is widespread
anticipation that detailed studies of $CP$ violation may lead us to
``New Physics" that goes beyond the Standard Model (SM).  In the
context of the SM, $CP$ violation is described by the phase in the
Cabibbo-Kobayashi-Maskawa quark-mixing matrix that is related to the
existence of six quark flavors grouped into 3 families.  $CP$
violation shows up in family-changing interactions, while in
family-conserving cases $CP$ violation is unobservably small.  The
last consequence needs experimental verification which is lacking thus
far.  Of the 30 tests of $CP$ listed in the Review of Particle
Physics~\cite{PDG01} only four are in this category: the $2\pi$ decays
of the $\eta$ and $\eta^{\prime}$ which at the present level are tests
of $P$ as well as $CP$.

Doable tests of $CP$ invariance are hard to find for lack of
particles or states which are eigenstates of the $CP$ operator
and have family-conserving interactions.  There are several
speculative ideas about unconventional $CP$ violation such as
spontaneous $CP$ violation in the extended Higgs sector but they
are hardly compelling.

\section{$\eta\to\pi\pi$}
The strong decay $\eta\to2\pi$ is forbidden by $CP$ and \emph{P}
invariance.  An $\eta$ can decay via the weak interaction; at the
level of $10^{-7}$, parity is not conserved any longer and for
$BR\lsim10^{-7}$, $\eta\to2\pi$ becomes a real test of $CP$
invariance.  There is considerable evidence that parity is conserved
in purely strong and electromagnetic interactions.  Strong-electroweak
interference can produce a violation of \emph{P} in carefully chosen
processes, usually at a small level which is of order $10^{-6}$. 

The current experimental limits on the $2\pi$ decay of the $\eta$ are:
\begin{equation}
BR(\eta\to\pi^0\pi^0) < 4.3\times 10^{-4}~\cite{PDG01}.
\end{equation}
\begin{equation}
BR(\eta\to\pi^+\pi^-) < 3.3\times 10^{-4}~\cite{PDG01}.
\end{equation}
Both limits have been obtained at a $\phi$ factory where $\eta$'s are
produced in the decay $\phi\to\eta\gamma$ ($BR = 1.2\%$).

There are no $\eta$ beams, as the $\eta$ lifetime is too short.  The
$\eta$'s come from baryon decays, in particular from the
$N(1535)\frac{1}{2}^-$ and $\Lambda(1670)\frac{1}{2}^-$ resonances,
and from meson decays, observed e.g.\ in $\phi$ decay or in $\overline{p}p$
annihilation.  In all cases there is plenty of $2\pi$ production; it
is comparable to $\eta$ production.  Thus, it is a real experimental
challenge to push the $\eta\to2\pi$ limit down below the $10^{-7}$
level. 

As indicated earlier, since $\eta\to2\pi$ is a flavor-conserving
interaction the expected $BR$ in the SM is small.  A recent
calculation yielded $BR(\eta\to2\pi)<3\times10^{-17}$~\cite{Sha01}.
The discovery of a much larger decay rate would be a sign for the
existence of a nonconventional $CP$ violating mechanism.

\section{$\eta\to4\pi^0$, a new test of $CP$}
The upper limit for the $CP$ test $\eta\to2\pi$ is hard to
improve appreciably with currently available setups because of the
sizeable $2\pi$ background in every $\eta$ production reaction.  It is
thus of interest to find another test.  A novel possibility is
$\eta\to4\pi^0$ which is forbidden by $CP$ and $P$.  For
$\eta$'s produced in the reaction $\pi^-\ p\to\eta n$ near threshold
there is no known background to $\eta\to4\pi^0$.  The chief drawback
is the smallness of the final state phase space for $\eta\to4\pi^0$
compared to $\eta\to2\pi^0$. 

The Crystal Ball Collaboration has recently produced the first upper
limit~\cite{Pra00a}:
\begin{equation}
BR(\eta\to4\pi^0) < 6.9 \times 10^{-7}.
\end{equation}
Together with Eq.~\ref{eq:etawidth}, this gives
$\Gamma(\eta\to4\pi^0)<8.9\times10^{-4}$ eV.  No events were found in
a sample of $3\times10^7$ $\eta$ decays produced near threshold in
$\pi^- p\to\eta n$ close to threshold.  To evaluate the sensitivity of
this test, note that the $\eta$ meson is an eigenstate of the $CP$
operator.  This allows for a comparison with a related but
$CP$-allowed decay.  The decay of a hypothetical $\eta$ meson, the
$\eta_{hyp}$, with $J^{PC}=0^{++}$ into $4\pi^0$ is allowed.  As
$\eta_{hyp}$ does not exist, we can instead use $f_0(1500)\to4\pi^0$.
The $f_0$ has the same quantum numbers as the $\eta$ except positive
parity.  The experimental value for the partial width is
$\Gamma(f_0\to4\pi^0) = 33$ MeV.  The ratio of the phase space
compared to $\eta\to4\pi^0$ is~\cite{Won00} $\Phi(\eta\to4\pi^0) /
\Phi(f_0\to4\pi^0)=$ $4.7\times 10^{-8}$, so we might expect
$\Gamma(\eta_{hyp}\to4\pi^0) \simeq 1.6\,\mathrm{eV}$.  Thus, the
$CP$-violating amplitude for $\eta\to4\pi^0$ compared to a comparable,
allowed decay is
\begin{equation}
A_{\overline{cp}} / A_{cp} < 
\left[
      \frac{8.9\times10^{-4}\,\mathrm{eV}}
           {1.6\,\mathrm{eV}}
\right]^{\frac{1}{2}} = 2.3\times 10^{-2}
\end{equation}
at 90\% CL.

\section{Charge Conjugation}
$C$ invariance, or charge conjugation symmetry, is the invariance of a
system to the interchange of the colored quarks with their antiquarks
of anticolor, the charged leptons with their antileptons, the
left(right)-handed neutrinos with the left(right)-handed
antineutrinos, and vice versa.  According to QED and QCD, $C$
invariance holds for all purely electromagnetic and all strong
interactions, but the experimental limits are not impressive.  The
Review of Particle Physics~\cite{PDG01} lists ``all weak and
electromagnetic decays whose observation would violate conservation
laws.''  Seventeen tests of $C$ invariance are listed: eight involve
decays of the $\eta$, six of the $\eta^\prime$, two of the $\omega$
and one of the $\pi^0$.  None has yielded a significant limit thus
far~\cite{Nef92}.  Neither has the Pais test~\cite{Pai59} which is
the equality of any pair of $C$-symmetric reactions.  Presented in
ratio form, there is $R_1 = \sigma(\overline{p} p\to
K^+X^-)/\sigma(\overline{p}p\to K^-X^+)$ and $R_2 =
\sigma(\overline{p}p\to \pi^+Y^-)/\sigma(\overline{p}p\to \pi^-Y^+)$.
The experimental data are $(R_1-1)< 2\times 10^{-2}$ and $(R_2-1)<
1\times 10^{-2}$~\cite{Bal65}, which are not sufficiently precise
for a useful analysis.  The paucity of tests of $C$-invariance is
related to the small number of eigenstates of the $C$-operator: only
flavorless mesons such as $\eta$, $\eta^\prime$, and self-conjugate
systems like $(p\overline{p}), (e^+e^-)$, and $(K^0\overline{K^0})$
qualify.

According to Stuckelberg and Feynman, an antiparticle may be viewed as
an ordinary particle that goes backward in time.  This shows that the
$C$, $P$, and $T$ symmetries are interwoven. $C$ reverses the sign of
all additive quantum numbers of a particle but leaves its spin
unaffected.  Thus the $C$ operator turns a left-handed neutrino into a
left-handed antineutrino.  The neutrinos studied in the lab all turn
out to be left-handed and the antineutrinos right-handed, which means
that there is full $C$ violation of the weak interactions.  The SM
does not explain this; it merely is part of the input of the SM,
namely it is assumed that all basic fermions come as left-handed
doublets and right-handed singlets.  This blatant asymmetry provides a
strong impetus for making better experimental tests of the validity of
$C$ invariance. 

Another argument which has kindled the interest in $C$ is the
experimental observations of the abundance of matter over antimatter
in the universe, as well as photons over baryons:
$n_B/n_\gamma<10^{-10}$~\cite{Coh93}.  In big-bang models of cosmology
one na\"{\i}vely expects the same abundance of matter and antimatter.
The known $CP$ violation is insufficient for explaining the
experimental baryon/antibaryon asymmetry.

Finally, recent neutrino experiments have provided tantalizing hints
of possible neutrino mixing and of a finite neutrino
masses~\cite{Fuk01,Ahm01}. 

\section{$\eta\to\pi^0\pi^0\gamma$, a new Test of $C$ Invariance}
The eta meson has the charge conjugation eigenvalue $C=+1$, and the
$\pi^0\pi^0\gamma$ system with $J^P=0^-$ has $C=-1$.  Thus, the
decay $\eta\to\pi^0\pi^0\gamma$ is strictly forbidden by $C$
invariance.  This decay would be an isoscalar electromagnetic
interaction of hadrons.  It has been suggested that there may exist an
isotensor electromagnetic interaction with a $C$-violating
component~\cite{Dom66,San71}.  The decay $\eta\to\pi^0\pi^0\gamma$
provides an opportunity to search for such an exotic interaction; it
would be a clear signal for New Physics. 

No searches for $\eta\to\pi^0\pi^0\gamma$ have been reported in the
literature.  A preliminary upper limit has been obtained using the
Crystal Ball detector~\cite{Pra01a} from a sample of $1.9 \times 10^7$
$\eta$'s.  Candidate events in the 
signal region are predominantly $(\sim 85\%)$ due to $\eta\to3\pi^0$
decay with overlapping photon showers;  the rest are due to $2\pi^0$
production with a split-off photon.  The net yield is no events
resulting in 
\begin{equation}
BR(\eta\to\pi^0\pi^0\gamma) < 
   5\times 10^{-4} \mathrm {\ at\ the\ 90\%\ C.L.}
\end{equation}
This corresponds to
$\Gamma(\eta\to\pi^0\pi^0\gamma)<0.6\,\mathrm{eV}$.  To evaluate the
sensitivity of this new result, we compare the upper limit with that
of a $C$-allowed decay.  In the absence of information on
$f_0\to\pi^0\pi^0\gamma$ --- the $f_0$ is the preferred comparison
state since it has $I^G(J^{PC}) = 0^+(0^{++})$ --- we use
$\rho\to\pi^+\pi^-\gamma$, with $BR = 1.0\times 10^{-2}$, $\Gamma_\rho
= 151\ \mathrm{MeV}$. In $\eta\to\pi^0\pi^0\gamma$ decay the $2\pi^0$
must be in a relative d-state.  This implies that the decay goes by a
magnetic quadrupole transition rather than a dipole as
$\rho\to\pi^+\pi^-\gamma$ does, and we must include a reduction factor
of order $(kR)^4$ (see Section~\ref{sec:4gamma}); we estimate this
factor to be $(\frac{1}{2})^4 = 0.06$ at worst.  After small
adjustments for the difference in phase space and the Clebsch-Gordan
factor, we find that a $C$-allowed decay has an expected decay width
of $2\times10^3$ eV.  This value is in reasonable agreement with two
other estimates; the first is based on the allowed decay $\rho\to2\pi$
and the other on the suppressed decay $\phi\to\pi^0\pi^0\gamma.$  The
sensitivity of the CB upper limit for $\eta\to\pi^0\pi^0\gamma$ is 
\begin{equation}
A_{\not{C}}/A_C < \left[\frac{0.6\,\mathrm{eV}}
                             {2\times10^3\,\mathrm{eV}}
                  \right]^{\frac{1}{2}} = 
   1.7\times 10^{-2}.
\end{equation}
We are not aware of a more precise test of $C$ invariance of an
isoscalar interaction.

\section{$\eta\to\pi^0\pi^0\pi^0\gamma,$ a test of C Invariance}
The radiative decay $\eta\to\pi^0\pi^0\pi^0\gamma,$ is strictly
forbidden by charge-conjugation invariance.  No search for it has been
published thus far.  There are seven photons in the final state, which
explains the need for a $4\pi$ acceptance detector.  The background is
mainly from $\eta\to3\pi^0$ with either a split-off or an old photon
shower from a previous interaction. 

Recently a preliminary result, an upper limit, has been obtained
using the Crystal Ball detector in an AGS experiment~\cite{Pra00}:
\begin{equation} 
BR(\eta\to\pi^0\pi^0\pi^0\gamma) < 7\times10^{-5},
\end{equation}
which corresponds to
\begin{equation}
\Gamma(\eta\to\pi^0\pi^0\pi^0\gamma) < 9.0\times10^{-2}\,
     \mathrm{eV,\ at \ the\ 90\% \ C.L.}
\end{equation}

This is a test of an isovector electromagnetic interaction of
hadrons.  The sensitivity of this test has been evaluated using a
similar approach as was used in the previous section.  Starting from
the strong decay of the $\omega$-meson, $J^P = 1^-$,
$\Gamma(\omega\to\pi^+\pi^-\pi^0) = 7.5$ MeV. We estimate
the unknown radiative decay to be $\alpha$ times the strong decay
width.  Including an adjustment factor for the difference in phase
space and the spin average weight factor~\cite{Gar01} we obtain for an
allowed $3\pi^0\gamma$ decay rate (if $C$ invariance did not
exist) $6.8\times10^3$ eV.  The upper limit for a $C$-violating
amplitude is thus 
\begin{equation}
A_{\not{C}}/A_C \leq
   \left[\frac{9\times 10^{-2}\,\mathrm{eV}}
              {6.8\times10^3\,\mathrm{eV}}
         \right]^{\frac{1}{2}} = 
   3.6\times10^{-3}.
\end{equation}
This is the best upper limit for an isovector electromagnetic
transition. 

\section{$\eta\to3\gamma$}
The decay of a neutral, flavorless, $C=+1$, pseudoscalar meson into
three photons is forbidden rigorously by $C$-invariance.  The
3$\gamma$ decay should be small as it is a third order electromagnetic
interaction and $\alpha^3=4\times10^{-7}$.  The rate is further
suppressed by substantial factors coming from phase space and angular
momentum barrier considerations~\cite{Ber65}.  The decay
$\eta\to3\gamma$ can be isoscalar or isovector and even the
hypothetical isotensor interaction.  The Particle Data
Group~\cite{PDG01} lists the upper limit for the $\eta\to3\gamma$
branching ratio as $5\times10^{-4}$. 

The Crystal Ball experiment at the AGS has produced a new, still
preliminary result which is~\cite{Pra00,Pra01b} 
\begin{equation}
\label{eq:3gammaexp}
BR (\eta \to 3 \gamma) < 1.8 \times 10^{-5}
\end{equation}
at the 90\% C.L.  Using Eq.~4, this corresponds to 
\begin{equation}
\Gamma (\eta \to 3\gamma) < 2.3 \times 10^{-2}\,\mathrm{eV}.
\end{equation}
The largest background in this experiment is from $\eta\to3\pi^0$
decay when photon showers overlap in the detector.  The background
from 
$\eta\to\pi^0\gamma\gamma$ with two overlapping photon showers is at
the percent level, because $BR(\eta\to\pi^0\gamma\gamma)$ is only
$3\times10^{-4}$ (see Section~\ref{sec:pi0gg}).  The background from
$\eta\to2\gamma$ with a split-off is totally suppressed in our
analysis. 

There is no straightforward way to assess the sensitivity of the $\eta
\to 3\gamma$ process analogous to the one used in the preceeding two
sections.  The triplet positronium state decays into $3 \gamma$ but it
has $J^{PC}=1^{--}$, which is not quite suitable for the task at
hand. 

The decay $\eta\to3\gamma$ can take place by the allowed,
$C$-violating, $CP$ conserving weak interaction of the SM denoted by
$BR (\eta \to 3 \gamma)_w$.  Using a quark-loop model
Dicus~\cite{Dic75} has obtained 
\begin{equation}
BR(\pi^0 \to 3\gamma)_w = 
     (1.2\times10^{-5}) 
     \frac{\alpha}{(2\pi)^5}G^2m^4_\pi\left(\frac{m_\pi}{m}\right)^8,
\end{equation}
where $G$ is the Fermi constant and $m$ is an effective quark mass.
Choosing $m > (1/7)m_N$ this yields
\begin{equation}
BR (\pi^0 \to 3\gamma)<6 \times10^{-19}.
\end{equation}
A similar expression for the allowed weak decay of the $\eta$ yields
for $m > (1/5)m_N$ the limit~\cite{Her88} 
\begin{equation}
BR (\eta\to 3\gamma)_{\omega} \ \ < 3 \times 10^{-12}.
\end{equation}
$\eta \to 3\gamma$ decay due to a $CP$-violating new interaction is not
likely.  P. Herczeg~\cite{Her87} has shown that in renormalizable gauge
models with elementary quarks the flavor conserving nonleptonic
interactions of the quarks do not contain in first order a
$P$-conserving $CP$-violating component.  The $CP$-violating contributions
to $BR(\eta \to 3 \gamma)$ in such models are therefore negliable
relative to $BR(\eta\to 3\gamma)_{\omega}$.    P. Herczeg~\cite{Her88} has
considered the existence of a flavor conserving $C$- and $CP$-violating
interaction $(\bar{H})$.  Using the stringent limits imposed by the
upper limit of an electric dipole moment of the neutron, he obtains
\begin{equation}
B (\eta \to 3\gamma)_{\bar{H}} < 10^{-19}.
\end{equation}

It is of interest to compare the relative sensitivity of the decay
rates of the three lightest 
pseudoscalar mesons into 3$\gamma$. The simplest effective
Hamiltonian for a $0^-$ meson decaying into 3$\gamma$ contains seven
derivatives~\cite{Her88}, consequently,
\begin{equation}
BR \sim m^{12}_{0^-} \cdot \Gamma(0^- \to all).
\end{equation}
This results in the following sensitivity comparison
\begin{equation}
BR(\eta \to 3\gamma) \ : \ BR(\pi^0 \to 3\gamma) \ : \ BR 
(\eta^{\prime} \to 3\gamma) = 1 : 10^{-5} \ : \ 5.
\end{equation}
Thus, even though the experimental limit for $BR(\pi^0 \to 3 \gamma)<3
\times 10^{-8}$ is small, it is 100 times less sensitive than
the experimental upper limit for $\eta \to 3 \gamma$ given in
Eq.~\ref{eq:3gammaexp}.  On the other hand, $BR(\eta^{\prime}\to 3
\gamma)<1 \times10^{-4}$ quoted in Ref.~\cite{PDG01} is a comparable
test of $C$.

In recent years there has been a growing interest in quantum field
theory over noncommutative spaces in part because of the connection to
string theories.  Noncommutative $CP$-violating effects are now being
estimated.  They may actually dominate over the SM
contribution~\cite{Hin01}. 

Grosse and Liao~\cite{Gro01} have shown that a generalization of the
anomalous $\pi^0\to2\gamma$ interaction can induce a $C$-violating
$\pi^0 \to 3\gamma$ amplitude in noncommutative quantum
electrodynamics.  The prediction for $BR (\pi^0 \to 3\gamma) = 6
\times 10^{-21}$ is still far from the experimentally reachable level
but it shows that there are new options.  It will be interesting to
see what noncommutative field theory will predict for $\eta \to
3\gamma$. 

\section{Weak Neutral $\eta$ Decays}
Most of $\eta$ physics is focused on the investigations of the strong
and electromagnetic interaction of confined $q\bar{q}$'s.  It is
appropriate to note that weak $\eta$ decay could occur already at
a BR level of $10^{-7}$.  The opportunity to actually observe a
``standard''  weak $\eta$ decay is rather slim because of various
suppression factors and the fact that the $\eta$ is a neutral meson
which eliminates charged weak leptonic decays.  As a result the
permitted weak $\eta$ decays in the standard model are expected to
occur at the level $10^{-13}$ and below~\cite{Sha01}.  This situation
makes rare $\eta$ decays a fair hunting ground for searching for new
interactions which are mediated by new particles such as leptoquarks
which are not subject to the constraints of the ordinary weak
interactions. 

Recent new experimental data from neutrino detectors, in particular
Super-Kamiokande~\cite{Fuk01} and SNO~\cite{Ahm01} are being interpreted as
evidence for neutrino oscillations.  One or more neutrinos are
expected to have a mass.  These new developments are enhancing the
interest in very rare $\eta$ decays which involve neutrinos.  Weak
semileptonic $\eta$ decays are forbidden by $G$-parity conservation.
They can occur by second class weak currents and are much
suppressed~\cite{Sha01}.  This leaves us with double neutrino decays
which experimentally look like $\eta\to\mathit{nothing}$.  They can only be
investigated indirectly.  The present upper limit for
$\eta\to\mathit{nothing}$ is a paltry 2.8\% at the 90\%
C.L.~\cite{Abe96}  The experiment used an enhanced $\eta$ source based
on the Jacobian-peak method and clean tagged $\eta$'s as well as the 
determination of all $\eta$ decay modes.  The limit deserves to be
improved using a good $4 \pi$-acceptance detector.

There are three categories of $\eta$ decay into two neutrinos.  The
first category consists of decays which are allowed by lepton number
conservation, $\nu_e\overline{\nu_e}$, $\nu_\mu\overline{\nu_\mu}$ and
$\nu_\tau\overline{\nu_\tau}$.  They can only occur if neutrino states
of both 
chiralities exist, and imply New Physics which goes beyond the
minimal SM.  When neutrinos have mass the branching ratio is~\cite{Her88}
\begin{equation}
\Gamma (\eta \to \nu \bar{\nu}) \simeq 4 \times 10^{-9} 
(m_\nu/m_\eta)^2.
\end{equation}
Even the $\tau$- neutrino which has the poorest limit on its
mass, $m_{\nu_\tau}<18$ MeV, still gives a discouragingly small $\eta$
decay branching ratio, $< 4 \times 10^{-12}$. 

The second category contains the lepton-family violating but
lepton-number conserving $\nu\overline{\nu}$ final states such as
$\nu_e\overline{\nu_\mu}$, $\overline{\nu_e}\nu_\mu$,
$\nu_\mu\overline{\nu_\tau}$, etc.  This is of special interest in
view of the new data on neutrino oscillations.  

In the third category are the decays which violate lepton number by
two, such as
$\eta$ decay into $\nu_e\nu_e$, $\nu_\mu\nu_e$, etc.  They may
be generated by certain classes of leptoquarks.  D. Wyler~\cite{Wyl90}
has discussed bounds on the masses of various types of leptoquarks, B-
photons, and other exotics, based on the upper limits for different
leptonic and 
semileptonic pseudoscalar meson decays.  He finds 
\begin{equation}
BR (\eta \to \nu \nu) \simeq 10^{-6} (G^2_{lq}/G^2_F)
\end{equation}
where $G_{lq}$ is the coupling strength of the leptoquark and $G_F$
the weak coupling strength.  With $G_F\simeq(300\,\mathrm{GeV})^{-2}$,
we can relate the upper limit of rare eta decays to the mass $M_{lq}$
and relevant coupling constant $G_{lq}$ for instance
\begin{equation}
M_{lq} > G_{lq}(300\,\mathrm{GeV})/33 [BR (\eta \to \nu
\bar{\nu})]^{1/4}.
\end{equation}

The helicity suppression which hampers $\eta \to \nu\overline{\nu}$
decays by the factor $(m_\nu/m_\eta)^2$ does not apply to $\eta \to
\nu \bar{\nu} \gamma$ because the spin of the photon allows the $\nu
\overline{\nu}$ pair to have angular momentum $J=1$.  The expected BR
is still very low.  Arnellos \emph{et al.}~\cite{Arn82} find using the
standard quark model 
\begin{equation}
BR (\eta \to \nu\overline{\nu} \gamma) \simeq 2 \times 10^{-15}.
\end{equation}

The weak decay $\eta\to\pi^0\nu\overline{\nu}$ has been discussed
briefly by P. Herczeg~\cite{Her88}.  In the SM it is a second class weak
interaction with $BR \simeq 10^{-13}$.  Beyond the SM $\eta \to \pi^0
\nu \bar{\nu}$ could proceed via an interaction which couples
neutrinos to a scalar quark current.  The branching ratio is not
expected to be larger than $10^{-9}$~\cite{Her88}.

The largest predicted weak decay of the $\eta'$ is $\eta'\to K\pi$,
which could be in the range $BR\simeq10^{-9}$~\cite{Ber90}.

\section{Summary}
The different neutral $\eta$ decay modes are listed in
Table~\ref{tab:alldecays} together with the experimental branching
ratio or upper limit as well as the chief physics interest.  The list
shows tests of $C$ and $CP$ invariance and chiral perturbation theory.
The rare and forbidden $\eta$ decays are useful for placing limits on
several proposed modifications of the Standard Model and on
manifestations of New Physics. 

\begin{table}
\caption{\label{tab:alldecays}\small The Neutral $\eta$ Decays.}
\begin{center}
\begin{tabular}{ccl}\hline\hline
Decay Mode   & Branching Ratio  & Physics Highlight \\
\hline
All Neutrals             & $(71.6\pm0.4)\%$                     \\
$2\gamma$                & $(39.3\pm0.3)\%$ & SU(3) octet-singlet mixing \\
$3\pi^0$                 & $(32.2\pm0.3)\%$ & $\chi PTh$; $m_u-m_d$     \\
$\pi^0\gamma\gamma$      & $(3.2\pm0.9)\times10^{-4}$ & $\chi PTh$,
$O(p^6)$ \\
$2\pi^0$                 & $<4.3\times10^{-4}$ & $P$ and $CP$ \\
$4\pi^0$                 & $<6.9\times10^{-7}$ & $P$ and $CP$ \\
$\pi^0\pi^0\gamma$       & $<5\times10^{-4}$   & $C$ (isoscalar) \\
$\pi^0\pi^0\pi^0\gamma$  & $<4.7\times10^{-5}$ & $C$ (isovector) \\
$3\gamma$                & $<4.5\times10^{-5}$ & $C$ (isovector, isoscalar) \\
$4\gamma$                & $<2.8\%$ & \\
$\pi^0\pi^0\gamma\gamma$ & $<3.1\times10^{-3}$ & $\chi PTh$, New Physics \\
$\nu_e\overline{\nu_e}$  & $<2.8\%$ & New Physics, leptoquarks \\
$\nu_e\bar{\nu_\mu}$     & $<2.8\%$ & New Physics, leptoquarks \\
$\nu_e\nu_e$             & $<2.8\%$ & New Physics, leptoquarks \\
$\gamma\nu\overline{\nu}$           & $<2.8\%$ & New Physics, leptoquarks \\
$\pi^0\nu\overline{\nu}$            & $<2.8\%$ & New Physics, leptoquarks \\
\hline\hline
\end{tabular}
\end{center}
\end{table}


\end{document}